# Control of Surgical Gesture under Lingual Electro-Tactile Stimulation

F. Robineau[a], F. Boy[b], J-P. Orliaguet[b], J. O. Vázquez-BuenosAires[a], J. Demongeot[a] & Y. Payan[a].

**Abstract** — Performing minimal-invasive surgical punctures require guiding a needle toward an intracorporeal clinically-defined target. As this technique does not involve cutting the body open, a visualization system is employed to provide the surgeon with indirect visual spatial information about the intracorporeal positions of the tool. One may consider that such systems reduce the ergonomics of the situation as they generate a decorrelation between the actual movement of the needle and the displayed information about this movement. The present study aims at assessing the guidance of an intracorporeal puncture under a lingual electrotactile sensory substitution device — the Tongue Display Unit (TDU) — with respect to the performance evidenced under a computer-rendered visualization system. The TDU device provides information about the deviation of the needle movement with regard to a pre-planned "optimal" trajectory. Experimental results show that - (1) the stimulations furnished by the TDU are efficient enough to guide accurately one's puncture, - and (2) that previous training using the visualization system does not improve performances obtained under subsequent TDU guidance. Finally, results are discussed in terms of efficiency and usability of the TDU in the context of guiding minimally-invasive surgical gesture.

## 1  Introduction

The development of Computer-Aided Surgical systems (CAS) to assist physicians in the realization of minimally-invasive (MI) intracorporeal therapeutic gestures is a growing domain (that aims at improving both the safety and the efficiency of the surgical procedures (Lavallée, Cinquin & Troccaz, 1997). The present exploratory research focuses on MI percutaneous abdominal puncture. This technique involves passing a needle to a specifically defined location in the patient's abdomen through trocars located into small incisions of the abdominal wall. The entire procedure is carried out by manipulating the needle outer extremity from outside the patient's body, the surgeon being deprived of direct visual feedback about the actual needle trajectory. Prior to the actual surgical procedure, images of the patient's abdomen (CT scans, MRI or echographies) are collected to allow the planning of the surgical gesture: entry and target points and trajectory between both points. During the surgical procedure the actual trajectory of the needle tip is compared to the planning thus providing the surgeon with information about the deviation of his/her gesture with respect to the "optimal planned trajectory". Different kinds of guiding systems have been proposed, most of them providing visual information (plotted onto a 2D screen) about the deviation between the actual and the planned trajectory (Dubois, Nigay, Troccaz, Carrat & Chavanon, 2001; Troccaz, Peshkin & Davies, 1997). This 2D display aims at allowing the surgeon to carry out the corrections needed to reach the target. However, such systems of sensory feedback impose a decorrelation between the actual movement and the visual information about this movement (i.e. incoherence between the plane of movement and the plane onto which the visual feedback is projected — Palluel-Germain, Boy, Orliaguet & Coello, 2005). The present study aims at evaluating an alternative answer to this drawback. The proposed solution consists in presenting the sensory feedback about the position of the surgical instrument with respect to the planned trajectory in another sensory modality that avoids such a decorrelation.

The pathfinding work of Bach-y-Rita and collaborators (Bach-y-Rita, Collins, Saunders, White & Scadden, 1969; Bach-y-Rita, 2003) evidenced that stimuli characteristics of one sensory modality (e.g. vision) could be transformed into stimulations of another sensory modality (e.g. touch). In the case of visuo-tactile substitution the so-called "sensory substitution" thus involves the conversion of light energy into mechanical energy.

[a] Technique in Imaging, Modeling and Cognition Laboratory (TIMC) – Institute for Applied Mathematics in Grenoble (IMAG) in the Institut d'Ingénierie de l'Information de Santé, Pavillon D, Faculté de Médecine, F-38706, La Tronche, Cedex, France.

[b] Perception-Action Team in the Laboratory of Psychology and NeuroCognition (LPNC) CNRS - UMR 5105, Pierre Mendès-France University, F-38042 Grenoble Cedex 9, France.

The first development of visuo-tactile substitution systems was designed to provide distal spatial information to blind people (Collins & Bach-y-Rita, 1973; Lenay, Gapenne, Hanneton, Marque & Genouëlle, 2003). The original TVSS — namely, Tactile Vision Substitution System — was composed of 400 stimulators (20×20 matrix, Ø 1mm each) placed on the chest or on the brow and rendered the images captured by a video camera into a "tactile image". Their results showed that blind individuals were capable to achieve 100% correct performance in an object recognition task after prior training of 50 trials. A number of converging experimental observations suggest that, after adequate training, subjects, using a TVSS fulfil a shape-recognition task, experience a "projection" of the objects that they perceived (in the tactile modality) in the external world (Bach-y-Rita, 1972). For instance, a sudden change in the zoom of the camera caused the participants to act as if they were approaching an obstacle (production of evasive behaviors).

Tackling the practical problems posed by the tactile stimulation of the skin, Bach-y-Rita and his collaborators turned to the electro-stimulation of the tongue surface (Bach-y-Rita, Kaczmarek, Tyler & Garcia-Lara, 1998). The human tongue seems indeed a highly dense (Trulsson and Essick, 1997), sensitive and discriminative (spatial threshold = 2 mm) matrix of tactile receptors that are similar to the ones of the skin. Moreover, the high conductivity offered by the saliva insures the electrical contact between the electrodes and the tongue surface. It is to note that these latter characteristics highlight (1) that the acuity of the tongue is greater than the one of the tip of the fingers; and (2) that the receptors located on the surface may be easily and selectively activated. Therefore, Bach-y-Rita and colleagues designed and evaluated a practical human-machine interface, namely the Tongue Display Unit (TDU). After adequate training with the TDU, blind subjects experienced the same "projection" of the objects (in the external environment) that are perceived in the tactile modality as the one experienced by subjects using a TVSS (Sampaio, Maris & Bach-y-Rita, 2001). This latter phenomenon was interpreted in terms of functional similarities between primary visual area (occipital cortex) and somatosensory cortex (parietal cortex) that allows the reception of afferent information originating in different sensory systems (Kandel, Schwartz & Jessel, 1991). Bach-y-Rita's TDU (1972) consists in a 2D array of micro-electrodes (12×12 matrix) held between the cheeks and positioned in close contact with the superior surface of the tongue. Each electrode can be individually activated, thus stimulating a restricted area on the tongue surface. The functioning of the stimulating matrix is controlled by an external electronic device interfaced with a computer.

The present study aims at evaluating the use of a simplified version of Bach-y-Rita's TDU, i.e. a 6×6 matrix, in the context of guiding minimally-invasive intracorporeal punctures. Subjects practised punctures on an artificial model of the human abdomen and were provided — via the TDU — with real-time information about the positioning of the needle with respect to a predefined optimal trajectory.

To this purpose, we compared the performances of naïve individuals placed in two conditions of movement guidance: either they were using a system of CAS intracorporeal guidance providing the operator with a 2D-visual information about the position of the needle with respect to the optimal trajectory or, they had to carry out the same task guiding the puncture with the information furnished by a simplified TDU device. It is to underline that in order to allow this comparison the only information conveyed by the system of visual guidance was reduced to a real-time display of the needle's orientation so that both systems may be considered as affording information of the same type. The present work attempts to address three questions: Is the TDU effective in guiding intracorporeal punctures? Is that guidance more efficient when individuals control the gesture through visual information than when they guide their movement through the information provided by the TDU? If it is the case, can the efficiency of the TDU be improved after a prior training under a system furnishing visual spatial information?

## 2 Experiment

### 2.1 Method

#### 2.1.1 Design

In this study we used an independent-measures design (between-subjects) in which two groups of individuals were required to carry out an intracorporeal puncture on an artificial model of the abdominal cavity. Subjects had to guide the surgical gesture in two conditions, either with the information displayed on a computer screen, or with the information furnished by a Tongue Display Unit (TDU).

Participants were required to carry out a minimal-invasive intra-corporeal puncture on a realistic 3D section of the abdomen ("Phantom", CIRS®, Norfolk, VA, USA) placed in ventral decubitus (Figure 1).

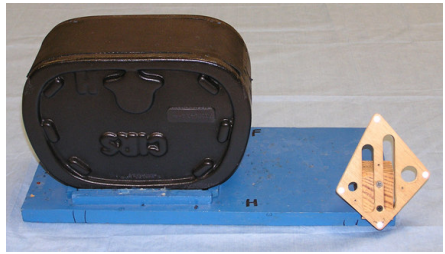

**Figure 1:** 3D abdominal model ("Phantom") used in the experiment (placed in ventral decubitus).

The experimental procedure was defined as follows: participants had to introduce progressively the surgical needle (length: 12cm, diameter: 1mm) until reaching an intracorporeal spherical target (diameter: 3mm). The target was located 95mm beneath the superior surface of the "Phantom". It is important to notice that the Phantom's both internal density and external texture resemble those observed in real human tissues. Though the 3D location of the target within the "Phantom" was identical for all trials, the external location of the entry point was randomly varied across trials —in order to avoid the production of stereotypic movements. Miniature infrared (IR) reflectors (Polaris®) were secured on the needle to allow determining both the needle's tip 3D position and bank angle (sampling frequency: 20Hz, spatial resolution: 0.2mm in every direction). Three additional IR reflectors were set on the Phantom so that to express the needle's positional data with respect to the target sphere within the faked abdomen (Phantom frame of reference). Dedicated software compared in real time the needle's actual path of movement with respect to an optimal trajectory defined as the straight line joining entry point to the target. We have to note that the target and therefore the optimal trajectory were defined prior to movement through 2D echographic imaging of the Phantom.

### 2.1.2 TDU System

The TDU consisted of a 2D array of 36 electrotactile electrodes (6×6 matrix, radius: .7mm each), embedded on a 1.5×1.5 cm plastic strip (Figure 2). The TDU was inserted into the oral cavity and maintained in close and permanent contact with the tongue all over the duration of the experiment (Figure 3). The matrix of electrodes was connected to an external electronic device triggering the electrical signals that stimulate the tactile receptors of the tongue (namely the Meissner corpuscle). As the saliva offers a good electrical conductivity the TDU only requires a 5 to 15 V output voltage and a 0.4 to 4 mA current to stimulate the tactile receptors in the tongue. Indeed, when one or more electrodes are activated, subjects feel a vibration on the surface of the tongue. In the present study we maintained constant the frequency of monophasic stimulation pulses (50Hz) across trials, subjects and conditions.

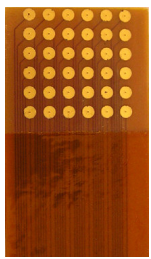

**Figure 2:** Tongue Display Unit. Each electrode measures 0,7mm in radius and is separated by a 0,9mm interval from its neighbours.

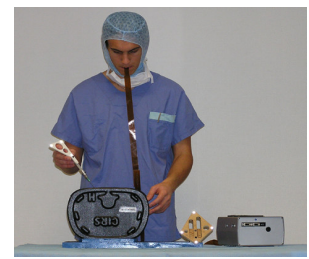

**Figure 3:** Subject guiding a puncturing needle under TDU guidance. The subject's gaze is directed at the patient.

The TDU system aims at transmitting sensory information to the tactile receptors of the tongue in order to provide subjects with information that allows the directional control of his/her percutaneous puncturing gesture.

When the position of the needle tip deviates from the predefined optimal trajectory the TDU system signals the inadequate position of the needle by activating several electrodes of the TDU matrix (see Figure 4). The activated zones on the matrix are symmetrical with respect to the direction of the inadequate needle bank. Eight different orientations were defined (N, S, E, W, NE, NW, SE, and SW) and a cross-stimulation indicated that the needle's tip overshot the plane of the target. No stimulation was sent as long as the actual movements of the needle did not deviate from the optimal trajectory of less than 3 mm (safety tunnel, diameter: 3mm). The intensity of the stimulation was regulated as a function of the norm of the deviation, i.e. the more the distance between the needle's end and the predefined trajectory augmented, the more the intensity of the stimulation increased. Moreover the lower tactile sensitivity that characterises the posterior region of the tongue (due to a lower density of tactile receptors) was compensated by doubling the intensity of the electrical stimulating current. In order to provide information about the relative distance between the needle's tip and the target, the temporal frequency of sound beeps was increased as the needle approached the target.

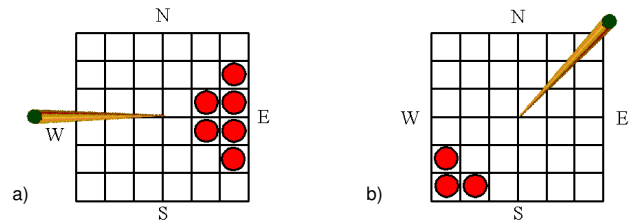

**Figure 4:** Examples of matrix coding and subsequent movement corrections (the needle is figured as an orange cone): (a) Subjects have to react to an eastern stimulation on the TDU by banking the needle on the opposite side (left inclination, west). (b) Similarly, correspondence for a south-western stimulation. Red dots symbolize activated unitary electrodes.

### 2.1.3 System used for visual guidance

The same task was used in the condition involving the visual guidance of the gesture (Figure 5). Computer-Assisted Medical Intervention software processed the coordinates of both the needle's tip and the target. The visual information available on a CRT video screen (21 inches, Sony®) provided a parsimonious two-dimensional view of the needle's tip and of the target (Figure 6). A grey immobile cross figured the direction of the optimal trajectory and a mobile red cross indicated the actual position of the needle's tip. This 2D view represented a plane centred on the target and perpendicular to the optimal predefined trajectory. Subjects had to maintain the needle within a zone of tolerance (equivalent to the one used in the TDU condition) while pulling the needle in the "Phantom". It is to note that a 1:10 scale ratio between the visual representation and the actual displacements of the needle allowed increasing the accuracy of the motor corrections. Similarly to the TDU condition, information about the relative distance between needle's tip and target was delivered through modulation of the temporal frequency of sound beeps.

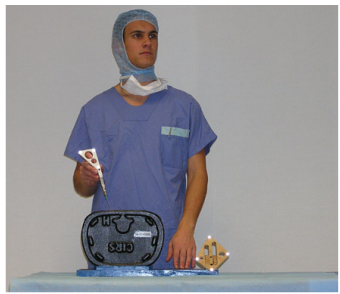

**Figure 5:** Subject guiding a puncturing needle under visual guidance: The subject looks at a screen display on a video monitor.

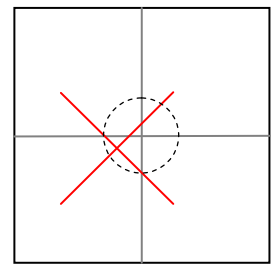

**Figure 6:** Schematic view of the screen display used in the condition involving visual guidance. This 2D visualization system figured the tip of the needle (red cross), the target (grey cross) and the orthogonal section of the "safety" tunnel (doted circle).

## 2.2 Procedure

Twelve students (18-25 years; 10 males & 2 females) and two surgeons participated in this experiment. They were naive as to the goals of the experiment. They had normal or corrected-to-normal vision and none of them reported suffering from troubles that could impair movement guidance.

### 2.2.1 Familiarization with the information afforded by the TDU

Prior to the experiment, participants were familiarized with electrotactile information provided by the TDU. This phase intended at determining individuals' optimal intensity threshold for the lingual electrotactile stimulation and explaining the associations between corrections in the orientation of the needle and the information provided by the TDU (Figure 4). Subjects were submitted to a training task (for 15mn) in which they had to discriminate and associate a stimulation (among the 8 azimuths the matrix could code) with a specific needle orientation. Thereafter subjects were submitted to the TDU condition.

### 2.2.2 Experimental Design

The goal of the present experiment was to compare the spatio-temporal characteristics of a minimal-invasive percutaneous puncture movement under two conditions of movement guidance, either via the tactile stimulation of the tongue (TDU condition) or, via a visual control (Visual control condition, V). The first group (TDU1 group) carried out one succession of 10 punctures using only the information provided by the TDU system (TDU condition), whereas the second (V1+TDU2 group) first accomplished the very same task using the visual guidance system (Visual control condition, V1 phase) and subsequently carried out 10 punctures using the TDU system (TDU2 phase). Thus we defined two conditions of TDU guidance: TDU1 and TDU2. It is to note that participants were instructed to reach the target at their best speed-accuracy trade off. No additional information about the spatial localization of the intracorporeal target or the adequate orientation of the needle was given. We excluded from the present analysis trials that lasted more than 200s (10% of the total 140 trials).

### 2.2.3 Parameters

The needle tip Cartesian coordinates (x, y, z) were processed under Matlab 6.5 (Mathworks®, USA). On the spatial side, for each trial, we also calculated the sum of the point-to-point vectors in order to assess the length of movement path (Movement Amplitude, MA). We computed a maximal distance in measuring — at each sampling interval — the distance between the actual position of the needle tip and the predefined "optimal" trajectory. The norm of the maximal deviation vector was taken as the maximal distance (Maximal Distance, MD). We also computed Total Movement Time (TMT).

## 3 Results

The first step of the result section aims at comparing the spatio-temporal parameters of movement in the TDU and V conditions. In other terms we intend at determining the basal level of performance of the sensorimotor apparatus when guiding the movement with each of the two systems. Analyses will involve comparing data collected in V1 and TDU1 phases. On the other side, our goal is also to determine whether the spatio-temporal performance of punctures carried out under the TDU condition (TDU2 phase) can benefit from a previous training in the visual modality (V1 phase). This new data analysis will involve comparing performances in both TDU1 and TDU2 phases. Six analyses of variance (mixed-design ANOVAs) will be processed in order to assess the evolution of each parameter along the ten trials of a movement production session (TDU1, TDU2 or V1 phases). Post hoc Fisher's Least Significant Differences (LSD) comparisons will be used to test for inter-trials differences (post-hoc pairwise comparisons).

### 3.1 Subjects' Successful Outcome in the TDU condition

In the TDU condition, 90% of the trials reached the target within 200s. Thus the present result shows that, after limited training (10 trials only), the guidance via lingual electrotactile stimulation appears to be efficient to control intracorporeal puncture gestures. We may suppose that the 100% baseline performance evidenced under the condition of visual guidance (i.e. V1) should be reached after sufficient training.

## 3.2 Spatial-Temporal Performance under TDU and Visual Guidance (TDU1 vs. V1)

### 3.2.1 Movement Amplitude (MA)

Figure 7 shows the mean MA for every trial for the TDU1 and V1 phases. An overall ANOVA on MA parameter shows a main effect of the conditions of movement guidance ($F(1, 7) = 7.2$, $p = .03$), movement amplitude being significantly lower under the condition of visual guidance (V1 phase) than under TDU guidance (TDU1 phase), no effect of the trial number ($F(9, 63) = 1.2$, $p = .32$), and no interaction between the condition of guidance and the trial number ($F(9, 63) < 1$, $p = ns$).

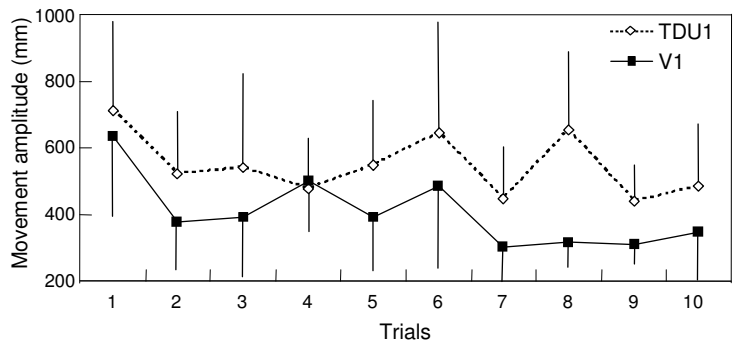

**Figure 7:** Evolution of movement amplitude (MA) as a function of the condition of movement guidance and of the trial.

### 3.2.2 Maximal Distance (MD)

Figure 8 shows the evolution of mean MD for every trial in both TDU1 and the V1 phases. Movements carried out under Visual control (V1 phase) tend to be less curved than movements carried out in the TDU condition ($F(1, 10) = 4.5$, $p = .06$). Therefore, under visual control, movement paths deviate less from the predefined "optimal" trajectory. In addition the ANOVA evidences no effect of the trial number ($F(9, 81) < 1$, $p = ns$) and no interaction between trial number and the condition of movement guidance ($F(9, 81) < 1$, $p = ns$).

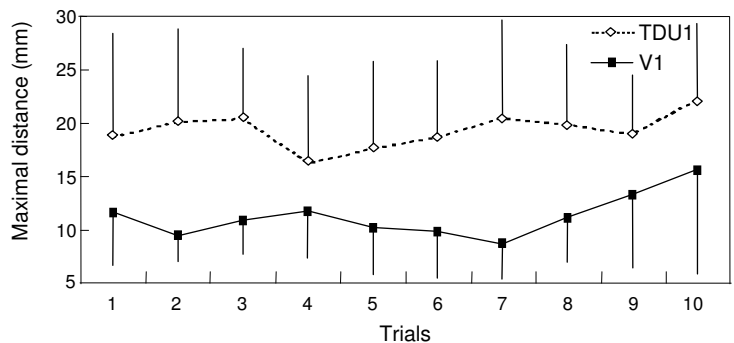

**Figure 8**: Evolution of maximal distance (MD) as a function of the condition of movement guidance and of the trial.

### 3.2.3 Total Movement Time (TMT)

Figure 9 shows the evolution of mean TMT for every trial for both TDU1 and V1 phases. The ANOVA shows no main effect of the conditions of movement guidance ($F(1; 10) = 2.4$, $p = .14$), an effect of the trial number

(F(9; 90) = 2.9, p = .005) and no interaction between trial number and condition of movement guidance (F(9; 90) < 1, p = ns). Thus, whatever the condition of movement guidance, the temporal performance cannot be considered as different.

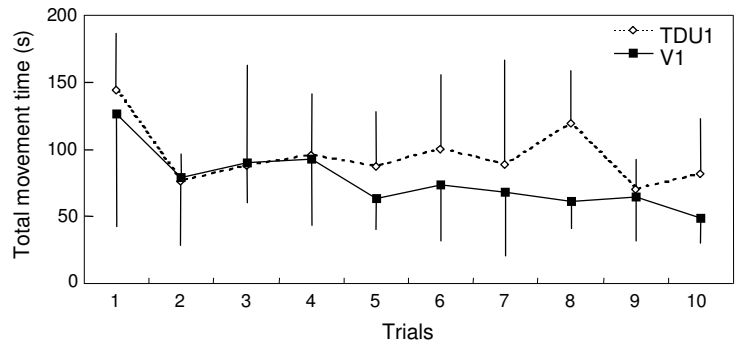

**Figure 9:** Evolution of total movement time (TMT) as a function of the condition of movement guidance and of the trial.

It follows from the above analyses that though temporal data (TMT) fails showing a difference between the two conditions of movement guidance, both spatial parameters (MD & MA) tend to evidence a superiority of visual guidance. This latter result is somewhat striking because though we were waiting for clear-cut differences between both conditions of guidance, subjects' performance appears only significantly different on the movement amplitude. Therefore, we may hypothesize that an improvement of performance in the TDU condition could appear after prior practice under visual guidance (intersensory transfer of motor learning). This later hypothesis will be tested through the comparison of performance collected in the independent TDU1 and TDU2 conditions of TDU guidance.

## 3.3 Spatial-Temporal Performance under TDU (TDU1 vs. TDU2)

### 3.3.1 *Movement Amplitude (MA)*

Figure 10 shows the mean MA for every trial in both TDU1 and TDU2 phases. The ANOVA shows -no difference between the different conditions of TDU guidance — i.e. no main effect — (F(1, 7) < 1, p = .4), -a significant main effect of the trial number (F(9, 63) = 6.2, p > .03) -and a non significant interaction effect (F(9, 63) = 2.2, p = .7). Post hoc inter-trials pairwise comparisons in the TDU1 phase evidenced a significant increase of movement amplitude between trial 7 & 8 (p = .008) and a significant reduction of MA between trials 8 & 9 (p = .02). In the TDU2 phase we noted a significant reduction of MA between trials 2 & 3 (p = .07). Post-hoc comparisons between conditions of TDU guidance (TDU1 vs. TDU2) for the same trial were all non significant (all p > .05).

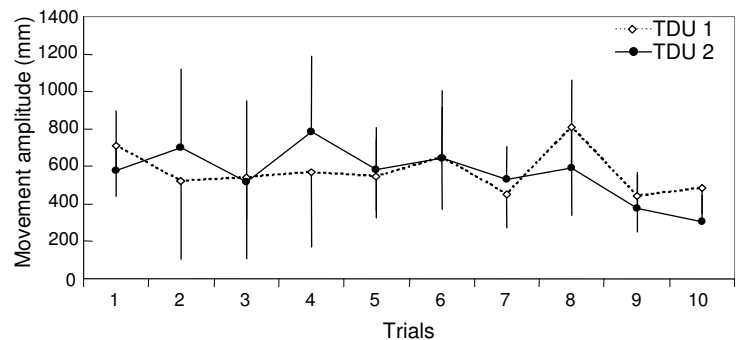

**Figure 10:** Evolution of movement amplitude (MA) as a function of the condition of movement guidance (TDU1 vs. TDU2) and of the trial.

### 3.3.2 Maximal distance (MD)

Figure 11 shows data concerning the MD parameter in the two conditions of TDU guidance (i.e. TDU1 & TDU2). The ANOVA shows: no difference between the different conditions of TDU guidance ($F(1, 7) < 1$, $p = .5$), a tendential effect of the trial number ($F(9, 63) = 1.93$, $p = .063$) and a tendential interaction effect ($F(9, 63) = 1.90$, $p = .067$). Post hoc comparisons evidence no inter-trial improvement in the TD1 condition. As concerns TDU2 intertrial differences were found significant between trials No. 2 & No. 3 ($p = .01$) and No. 8 & No. 9 ($p = .01$). Post hoc comparisons between mean MD values obtained in the two conditions of TDU guidance (TDU1 vs. TDU2) for the same trial lead to non significant differences (all $p > .05$).

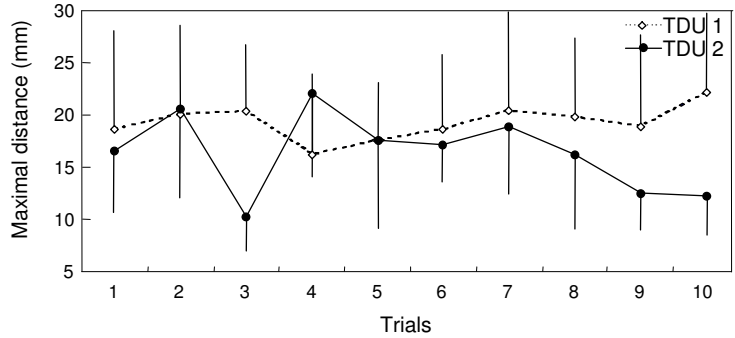

**Figure 11:** Evolution of maximal distance (MD) as a function of the condition of TDU guidance and of the trial.

### 3.3.3 Total Movement Time (TMT)

Figure 12 shows the evolution of mean TMT as function of both the condition of TDU guidance (TDU1 vs. TDU2) and of the trial number. The ANOVA shows no main effect of the conditions of TDU guidance ($F(1; 5) < 1$, $p = .9$), no effect of the trial number ($F(9; 45) = 1.3$, $p = .25$) and no interaction between trial number and condition of TDU guidance ($F(9; 45) < 1$, $p = .6$). Thus, whatever the condition of TDU guidance, temporal performances cannot be considered as different. Post hoc comparisons show a tendential intertrial improvement in the TDU1 condition for the 7th and 8th trials ($p = .09$). Inter-trials pairwise comparisons in the TDU2 phase evidenced a tendential improvement between trials 3 & 4 ($p = .09$). Comparisons between conditions of TDU guidance (TDU1 vs. TDU2) for the same trial were all non significant (all $p > .05$).

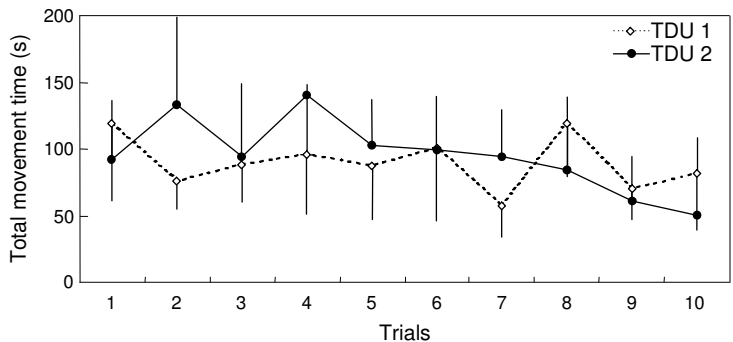

**Figure 12**: Evolution of total movement time (TMT) as a function of the condition of TDU guidance and of the trial.

The present results show that neither spatial (Movement Amplitude, Maximal Distance) nor temporal (Total movement time) parameters evidence a better performance in the TDU2 condition with respect to the TDU1 condition. Thus TDU guidance of movement (comparison between TDU1 & TDU2) is not improved by prior training under visual feedback (in V1 condition).

# 4    Discussion

The present research aimed at evaluating the use of a Tongue Display Unit in the context of guiding intracorporeal percutaneous punctures. The results are very encouraging for further development of electrotactile movement guidance devices in the domain of minimally-invasive surgical techniques.
On a general basis, the results highlight the fact that electrotactile stimulation of the tongue to guide one's movement appears to be efficient. Under TDU (TDU1 & TDU2 conditions) subjects indeed reached the intracorporeal target in following an almost direct and straightforward trajectory in 57% of the trials. In the remaining trials, the target was indirectly reached after prior overshooting.
When comparing performances in conditions of visual and TDU movement guidance (V1 vs. TDU1), results in the last trials of the experimental session (all post hoc comparisons > .05) can be considered as similar whatever the parameter (TMT, MA or MD). This result is somewhat surprising because there are evidences demonstrating that evaluation and reproduction of orientations is far more efficient in the visual than in the tactile or haptic modalities (Gentaz, Luyat, Cian, Hatwell, Barraud, & Raphel, 2001). On the contrary we demonstrated that, at least as concerns orientation coding, the TDU system allows a performance equivalent to the one observed under a simplified visual navigation system.
Moreover, we showed that the efficiency of TDU guidance was not improved by prior training under the visual guidance condition (TDU1 vs. TDU2 comparison). This latter result can be interpreted as denoting a difference between the processing of sensory information by the tactile and visual modality. This interpretation is backed up by several subjects' remarks stating that the use of the TDU system appeared more "intuitive" than the visual system. The subjects found the TDU system of guidance easy to use because it involved less cognitive activity such as interpreting the angular deviation on the visual display. The present results indeed confirm that the "costless and intuitive" way of coding angular deviation proposed by the TDU is efficient enough to afford the necessary information for accurate guidance of target-directed movement. Under TDU guidance, correcting one's puncturing movement thus involved applying to the needle an inclination in the opposite side (symmetrically) of the tongue stimulation.
Nevertheless one may think that anatomical, physiological and ergonomic constraints can hamper the usability of TDU devices. On the anatomical side, the within- and between-subject(s) heterogeneous distribution of tactile receptors on the surface of the tongue (lower density in the posterior region) has to be taken into account in regulating the intensity of the stimulation. The present study shows that doubling the intensity for Northern stimulating electrodes (N, NE & NW) was proved to be not sufficient enough as 64% of the stimulations were located in the part of the matrix placed upon the less discriminative part of the tongue (postero-superior area). Moreover increasing the intensity of the stimulating current -1) maximizes problems posed by repeated stimulations of the same tactile receptors (extinction of the discriminative power due to saturation of the receptor) and, -2) augments the diffusion of the stimulation by the saliva in the neighbourhood of a given electrode (reduction of the discriminative power). Further developments of TDU devices shall take into account these two phenomena. The present results reveal the usability of the TDU system in a surgical task. In an attempt to optimize the TDU device, we are currently developing and testing a wireless radio-controlled version of the 6×6 TDU matrix. This latter version will be inserted into a palatal prosthesis that aims at improving the ergonomics of the TDU in reducing the secretion of saliva. The wireless design is a particularly complex task that is going to be developed by the company Coronis-Systems (www.coronis-systems.com) that will provide *Wavenis*, its wireless Ultra Low Power technology which has been preferred to current radio frequency (RF) standards that all have shortcomings when considering Ultra Low Power challenges.
Though the use of a Tongue Display Unit does not bring a global solution to the navigation of instrument in the many mini-invasive surgical procedures (inability to code for distance and target location), it has to be stressed that the tactile stimulation of the tongue provide accurate and costless orientation information that can be used with only little prior training. Thus it may be suggested that a TDU — dispensing orientation information via a non visual channel — could be used in an attempt to liberate the surgeon's visual resources so that she/he could concentrate on other aspects of her/his own gesture.